\documentclass[preprint,aps,prd,showpacs,superscriptaddress,nofootinbib,tightenlines]{revtex4}

\usepackage{amsfonts}
\usepackage{multirow}
\usepackage{mathrsfs}
\usepackage{graphicx}
\usepackage{amsmath}
\usepackage{amssymb}
\usepackage{bm}
\usepackage{bbm}
\usepackage{color}
\usepackage{xcolor}
\usepackage{float}
\usepackage{slashed}

\def\OMIT#1{}

\newcommand{\nn}{\nonumber}

\newcommand{\beq}{\begin{equation}}
\newcommand{\eeq}{\end{equation}}
\newcommand{\bqa}{\begin{eqnarray}}
\newcommand{\eqa}{\end{eqnarray}}
\newcommand{\bseq}{\begin{subequations}}
\newcommand{\eseq}{\end{subequations}}

\begin{document}
\title{\mbox{}\\[10pt]
A factorization theorem connecting the light-cone distribution amplitudes of
heavy-flavor mesons in QCD and HQET}

\author{Saadi Ishaq~\footnote{saadi@ihep.ac.cn}}
\affiliation{Institute of High Energy Physics, Chinese Academy of Sciences, Beijing 100049, China}
\affiliation{School of Physical Sciences, University of Chinese Academy of Sciences,
Beijing 100049, China}

\author{Yu Jia~\footnote{jiay@ihep.ac.cn}}
\affiliation{Institute of High Energy Physics, Chinese Academy of
Sciences, Beijing 100049, China}
\affiliation{School of Physical Sciences, University of Chinese Academy of Sciences,
Beijing 100049, China}

\author{Xiaonu Xiong~\footnote{xnxiong@csu.edu.cn}}
\affiliation{School of Physics and Electronics, Central South University, Changsha 418003, China}
\affiliation{Institute for Advanced Simulation,
Institut f\"ur Kernphysik and J\"ulich Center for Hadron Physics,
Forschungszentrum J\"ulich, D-52425 J\"ulich, Germany}

\author{De-Shan Yang~\footnote{yangds@ucas.ac.cn}}
\affiliation{School of Physical Sciences, University of Chinese Academy of Sciences,
Beijing 100049, China}

\date{\today}
\begin{abstract}
The light-cone distribution amplitude (LCDA) of a heavy-light meson defined in
heavy quark effective theory (HQET), is a fundamental nonperturbative input
to account for innumerable $B$ meson exclusive decay and production processes.
On the other hand, the conventional heavy-flavored meson LCDA defined in QCD,
also ubiquitously enters the factorization formula for hard exclusive $B$ production processes.
Inspired by the observation that these two LCDAs exhibit the identical infrared behaviors,
yet differ in the ultraviolet scale of order $m_b$ or greater,
we propose a novel factorization theorem for the heavy-light mesons,
that the LCDA defined in QCD can be further expressed as a convolution between the
LCDA in HQET and a perturbatively calculable coefficient function thanks to asymptotic freedom.
This refactorization program can be invoked to fully disentangle the effects from three
disparate scales $Q$, $m_b$ and $\Lambda_{\rm QCD}$ for a hard exclusive
$B$ production process, particularly to facilitate the resummation of
logarithms of type $\ln Q/m_b$ and $\ln m_b/\Lambda_{\rm QCD}$ in a systematic fashion.
\end{abstract}


\maketitle

The hard exclusive hadron production is one of the major battlefields of perturbative QCD~\cite{Brodsky:1989pv}.
If one denotes the hard momentum transfer scale by $Q$,
the famous collinear factorization theorem~\cite{Lepage:1980fj,Chernyak:1983ej}
demands that the reaction amplitude involving a single hadron
can be expressed as the following convolution integral:
\beq
{\cal M} = \int_0^1\!\! dx\, T(x;\mu_Q) \, \Phi^{\rm QCD}(x;\mu_Q)+{\cal O}(1/Q),
\label{QCD:factorization:theorem}
\eeq
up to higher-twist corrections.
Here $0\le x \le 1$ signifies the light-cone momentum fraction of the quark inside the hadron.
$T$ represents the perturbatively calculable hard-scattering kernel, $\Phi^{\rm QCD}$ denotes the
nonperturbative, yet, universal,
leading-twist light-cone distribution amplitude (LCDA) of the hadron defined in QCD.
The factorization scale $\mu_Q$ lies between $Q$ and $M \sim \Lambda_{\rm QCD}$ ($M$ denotes the hadron mass), which
enters both $T$ and $\Phi$ in a prescribed manner such that the physical amplitude becomes
independent of this artificial scale. Specifically, the $\mu$ dependence of the QCD LCDA
is governed by a celebrated renormalization group equation,
usually referred to as the Efremov-Radyushkin-Brodsky-Lepage (ERBL) equation~\cite{Lepage:1979zb,Efremov:1979qk}:
\bqa
\mu_Q \frac{d}{d \mu_Q} \Phi^{\rm QCD}(x;\mu_Q)= {\alpha_s C_F \over \pi}
\int_0^1 \!\! dy\, V_0(x,y)\,\Phi^{\rm QCD}(y;\mu_Q),
\label{ERBL:evolution:eq}
\eqa
with the color factor $C_F={N_c^2-1\over 2 N_c}$, and $N_c=3$ is the number of color.
The evolution kernel for a helicity-zero meson reads
\beq
V_0(x,y) =
\left[{x\over y}\left(1+{1\over y-x}\right) \theta(y-x)+\left(\begin{array}{c}{x \rightarrow \overline{x}}
\\
{y \rightarrow \overline{y}}\end{array}\right)\right]_+,
\label{ERBL:kernel}
\eeq
with $\bar{x}\equiv 1-x$. This equation can facilitate
to resum large collinear logarithm of type $\alpha_s \ln {Q\over \Lambda_{\rm QCD}}$
in a typical hard exclusive reaction. Note the formalism in \eqref{QCD:factorization:theorem} applies to
any specifies of hadrons, irrespective of being light or heavy, provided that
$Q\gg M \geq \Lambda_{\rm QCD}$.

Alternative factorization framework also exists if the reaction involves a heavy-flavor meson, such as
$B$ exclusive decay process,  exemplified by
$B\to\gamma \ell \nu$~\cite{KPY,DS03,proofsBgamma}.
By exploiting the hierarchy $m_b\gg \Lambda_{\rm QCD}$, the HQET factorization theorem~\cite{Beneke:1999br,Beneke:2000ry}
demands that the $B$ exclusive decay amplitude
may also be cast into a convolution form:
\beq
{\cal M} = \int_0^\infty \!\! d\omega\, {\mathcal T} (\omega, Q, m_b; \mu_H) \, \Phi_+^{\rm HQET}(\omega;\mu_H)+{\cal O}(1/m_b),
\label{HQET:factorization:theorem}
\eeq
in the $m_b\to \infty$ limit.
Here $\omega$ signifies the light-cone momentum of the light spectator quark inside the $B$ meson.
${\mathcal T}$ again represents the perturbatively calculable hard-scattering kernel thanks to asymptotic freedom,
$\Phi_+^{\rm HQET}$ denotes the nonperturbative, yet, universal,
leading-power LCDA of the $B$ meson, with the $b$-quark field defined in the
heavy-quark-effective-theory (HQET)~\cite{Grozin:1996pq,Beneke:2000wa}.
For a discussion about the model-independent properties of the $B$ meson LCDA in HQET, see \cite{Lee:2005gza}.
A peculiar feature of $\Phi_+^{\rm HQET}(\omega)$ is that its positive Mellin moments become UV divergent,
where one usually imposes a UV cutoff $\Lambda$ to regularize.
The $m_b$ dependence is entirely encoded in ${\cal T}$ but not in $\Phi_+^{\rm HQET}$.
Unlike collinear factorization, here one does not distinguish the scales $Q$ and $m_b$ in the hard-scattering kernel.
The dependence of $\mathcal T$ and $\Phi_+^{\rm HQET}$ on the factorization scale $\mu_H$,
which lies between $m_b$ and $\Lambda_{\rm QCD}$, conspires to
counterbalance each other in the physical amplitude.

The scale dependence of the $B$ meson LCDA defined in the HQET side
is controlled by the famous Lange-Neubert evolution equation~\cite{Lange:2003ff}:
\beq
 \mu_H { d \over d \mu_H} \phi^{\rm HQET}_+( \omega;\mu_H) =
 -{\alpha_s C_F \over 4\pi}  \int_0^\infty \!\! d\omega^\prime \,
 \gamma_+ (\omega,\omega^\prime;\mu_H) \, \phi^{\rm HQET}_+(\omega^\prime;\mu_H),
\label{LN:evolution:eq}
\eeq
where the one-loop anomalous dimension $\gamma_+$ reads
\beq
\gamma_+(\omega,\omega^\prime;\mu_H) =
\left(\Gamma_{\rm cusp}^{(1)}\ln{\mu_H \over \omega}-2\right) \delta(\omega-\omega^\prime)
-\Gamma_{\rm cusp}^{(1)} \omega \left[{\theta(\omega^\prime-\omega)\over \omega^\prime(\omega^\prime-\omega)}+
{\theta(\omega-\omega^\prime)\over \omega(\omega-\omega^\prime)} \right]_+,
\label{LN:kernel}
\eeq
with $\Gamma_{\rm cusp}^{(1)}=4$. The explicit occurrence of $\ln \mu_H$ in the
anomalous dimension might look peculiar.
This evolution equation can be employed to resum large soft logarithms
of form $\alpha_s \ln^n {m_b\over \Lambda_{\rm QCD}}$ ($n=1,2$).

Apart from numerous $B$ exclusive decay modes, we emphasize that the HQET factorization formalism
\eqref{HQET:factorization:theorem} can also be fruitfully
applied to exclusive $B$ {\it production} processes.
Various exclusive $D^\pm$ production processes have been investigated long ago at tree level
essentially following the ansatz of \eqref{HQET:factorization:theorem},
yet coined with a different terminology -- heavy-quark recombination mechanism~\cite{Braaten:2001bf}.
It is worth noting that HQET factorization framework
presents a successful and economic account
for the $D^+/D^-$ production asymmetry observed in various
Fermilab fixed target experiments~\cite{Braaten:2001uu,Braaten:2002yt}.
Very recently, the exclusive processes $W\to B(D_s)+\gamma$ have been calculated to order $\alpha_s$
in the context of HQET factorization~\cite{Ishaq:2019zki}.
We note that some exclusive channels of $W$, $Z$
radiative decays into heavy-flavor mesons have previously been investigated
in the standard QCD collinear factorization~\cite{Grossmann:2015lea}.

One interesting question may be naturally posed: for a hard exclusive $B$ production process
with scale hierarchy $Q\gg m_b \gg \Lambda_{\rm QCD}$,
since both light-cone factorization \eqref{QCD:factorization:theorem}
and HQET factorization \eqref{HQET:factorization:theorem} appear to be applicable,
so how can one manage to make the most optimized predictions?

Surely both factorization approaches are based upon solid theoretical ground, nevertheless
each of which has its own strength and weakness.
As mentioned before, a notable merit of the light-cone approach is that
large collinear logarithm $\ln Q/m_b$ can be efficiently resummed, by considering
$m_b$ as an IR scale. However, an apparent shortcoming of this approach is that the
characteristic feature of the heavy-flavor meson is not
adequately utilized, and the phenomenological constraints on $B$ meson LCDA defined in QCD
are also limited.
From theoretical angle, it is evident that the QCD LCDA cannot be entirely nonperturbative,
since it entails the hard scale $m_b$, and it is definitely desirable if
this perturbative effect can be explicitly separated from the $B$ meson QCD LCDA.
On the other hand, the strength of the HQET factorization is that the heavy-quark
nature of the $B$ meson has been fully exploited, by treating $m_b$ as a UV scale.
Moreover, much phenomenological knowledge on the $B$ meson LCDA defined in HQET has been gleaned based
on intensive investigations on numerous $B$ decay processes over the past two decades.
The weakness of this approach for exclusive $B$ production is that the hard-scattering kernel
involves two disparate scales $Q$ and $m_b$, and the large collinear logarithm may potentially
ruin the convergence of perturbative expansion.

The goal of this paper is to show that these two factorization approaches can be fruitfully combined
to make optimized predictions for hard exclusive heavy meson production.
The key is to establish a factorization formula connecting two types of $B$ meson LCDAs defined
in both QCD and HQET.
As we shall see, through the refactorization program, the effects from three disparate scales,
$Q$, $m_b$ and $\Lambda_{\rm QCD}$, can be fully disentangled.

We first recapitulate how the $B$ meson LCDAs are framed in QCD and HQET.
For simplicity, we will concentrate on the $\overline{B}$ meson composed of a $b$ quark and
a light spectator antiquark $\bar{q}$.
Let a time-like four-vector $v^\mu$ represent the four-velocity of the $\overline{B}$ meson,
which satisfies $P^\mu=m_B v^\mu$ and $v^2=1$.
For convenience we also introduce a reference null vector $n^\mu$ satisfying $n^2=0$.
The explicit definitions for both LCDAs then become
\bseq
\begin{align}
& \Phi^{\rm QCD}(x,\mu_Q) \equiv f_{B}  \phi^{\rm QCD}(x,\mu_Q)=
-i\int
{d z^- \over 2\pi}\, e^{ixP^+ z^-} \langle 0|\bar
{q}(z)[z,0]\slashed n\gamma_5 b(0)|\overline{B}(P)\rangle,
\label{B:QCD:LCDA:def}
\\
& \Phi_+^{\rm HQET}(\omega,\mu_H)\equiv \hat{f}_{B} \phi_+^{\rm HQET}(\omega,\mu_H)=
\frac{-i}{m_B v^+}\int{dt\over 2\pi}\, e^{i\omega t}\langle 0|\bar
{q}(z)[z,0]\slashed n \gamma_5 h_{v}(0)|\overline{B}(v)\rangle,
\label{B:HQET:LCDA:def}
\end{align}
\label{Two:LCDAs:definitions}
\eseq
where $z^\mu=z^- n^\mu$ is also a null coordinate vector, $t=v\cdot z$.
We have also defined $P^+=n\cdot P$ and $v^+=n\cdot v$.
$[z,0]$ represents the light-like gauge link, inserted to ensure gauge invariance.
$x$ in \eqref{B:QCD:LCDA:def} designates the light-cone momentum fraction of the light spectator quark $\bar{q}$.
The QCD decay constant $f_{B}$ can be factored onto the HQET decay constant $\hat{f}_{B}$ by integrating
out hard quantum fluctuation of order $m_b$, through the
perturbative matching~\cite{Eichten:1989zv,Neubert:1993mb}:
\begin{align}
f_{B} =  \hat{f}_B (\mu_H) \left[ 1 - {\alpha_s C_F\over 4 \pi}
\left(3\ln {\mu_H \over m_b} + 2\right) \right] +\mathcal{O}\left(\alpha_s^2\right).
\label{hat:f:matching:fB}
\end{align}

Notice the only difference of two LCDAs in \eqref{Two:LCDAs:definitions} is
that the $b$ quark field is defined in QCD for the former, while defined in HQET for the latter.
Obviously these two LCDAs have drastically different ultraviolet behavior,
as is evident in the completely different evolution equations
\eqref{ERBL:evolution:eq} and \eqref{LN:evolution:eq}.
Nevertheless, it is crucial to observe that these two objects possess exactly the identical infrared
behavior, since HQET faithfully reproduces the IR aspects of QCD.
Since the perturbative scale $m_b$ is still encompassed in $\phi^{\rm QCD}$,
it sounds appealing to explicitly factor this short-distance effect out of the QCD LCDA.
Conceivably, this scale separation can be achieved through the following refactorization program:
\beq
\Phi^{\rm QCD}(x,\mu_Q) = \int_0^\infty\!\! d\omega\,Z(x,\omega, m_b;\mu_Q, \mu_H) \, \Phi^{\rm HQET}_+ (\omega,\mu_H),
\label{Our:factorization:theorem}
\eeq
The coefficient function $Z$ captures all the effect of order $m_b\gg \Lambda_{\rm QCD}$,
thus can be computed in perturbation theory
thanks to asymptotic freedom.
It can be organized as
\beq
Z(x,\omega, m_b;\mu_Q, \mu_H)= Z^{(0)}(x,\omega, m_b)+ {\alpha_s C_F\over 4\pi} Z^{(1)}(x,\omega, m_b;\mu_Q, \mu_H)+{\cal O}(\alpha_s^2).
\eeq

The physical picture underlying \eqref{Our:factorization:theorem} may also be lucidly envisioned
in the context of strategy of region~\cite{Beneke:1997zp}. When computing the perturbative correction
to the QCD LCDA in \eqref{B:QCD:LCDA:def}, the loop momentum flowing into the $b$ can be partitioned into
either {\it hard} ($l^\mu\sim m_b$) or {\it soft} ($l^\mu\ll m_b$) regions.
It is the soft region that is exactly responsible for the contribution to
the HQET LCDA defined in \eqref{B:HQET:LCDA:def}, which is also equivalent to taking $m_b\to\infty$ limit
prior to conducting the loop integration. Therefore, the $Z$ function in \eqref{Our:factorization:theorem},
which just accounts for the difference between these two LCDA, receives contribution solely from the hard loop region,
therefore can be accessed in perturbation theory thanks to the asymptotic freedom.

Before proceeding, we pause to remark that the refactorization program here is in spirit
analogous to the factorization of the LCDA of the doubly-flavored heavy quarkonium,
$B_c$, into the nonperturbative local NRQCD matrix element
multiplied with the perturbatively calculable coefficient function~\cite{Ma:2006hc,Xu:2016dgp}.
It is amusing to point out that, the structure of \eqref{Our:factorization:theorem}
also looks similar to the factorization formula that links the quasi and light-cone parton distributions,
which has been recently formulated in the context of the large-momentum effective theory by Ji~\cite{Ji:2013dva}.
There the light-cone distribution function has the support $(0,1)$, while
the quasi-distribution has unbounded support $(-\infty,+\infty)$.
Of course, a notable difference between these two situations
is that both $B$ meson distribution amplitudes in our case
are light-cone correlators.

Determination of the $Z$ function can be best fulfilled via the
standard perturbative matching procedure. Since the $Z$ factor is insensitive to the IR physics,
one can freely replace the nonperturbative $\overline{B}$ meson by
a fictitious one, {\it i.e.}, a free $b\bar{q}$ pair, and compute the corresponding
$\Phi^{\rm QCD}$ and $\Phi^{\rm HQET}_+$ in perturbation theory:
\bseq
\begin{align}
& \Phi^{\rm QCD}(x,\mu_Q)= \Phi^{\rm QCD\:(0)}(x,\mu_Q)+\frac{\alpha_s C_F}{4\pi}
\Phi^{\rm QCD\:(1)}(x,\mu_Q)+{\cal O}(\alpha_s^2),
\label{QCD:LCDA:series:expansion}
\\
& \Phi_+^{\rm HQET}(\omega,\mu_H) = \Phi_+^{\rm HQET\:(0)}(\omega,\mu_H)+\frac{\alpha_s C_F}{4\pi}
\Phi_+^{\rm HQET\:(1)}(\omega,\mu_H)+{\cal O}(\alpha_s^2).
\label{HQET:LCDA:series:expansion}
\end{align}
\label{Two:LCDAs:series:exp}
\eseq
One is then able to solve \eqref{Our:factorization:theorem} to deduce the $Z$ factor iteratively,
order by order in $\alpha_s$.
Note this matching procedure is similar to deducing the perturbative $Z$ factor
that connects the quasi parton distribution functions and light-cone parton distributions~\cite{Xiong:2013bka}.

One can set up the kinematic configuration for $|b\bar{q}\rangle$ at his disposal.
For example, one may simply follow \cite{Ishaq:2019zki} to choose a static $b$ and a moving $\bar{q}$.
Fortunately, by modeling the $B$ meson as a free $b \bar{q}$ pair with vanishing relative motion,
Bell and Feldmann had already computed $\Phi^{\rm QCD}$ and $\Phi^{\rm HQET}_+$ through
order $\alpha_s$ a decade ago~\cite{Bell:2008er}. Consequently, based on their results,
we can directly extract the intended order-$\alpha_s$ part of the $Z$ function.
There the spectator light quark $\bar{q}$ is endowed with a non-vanishing constitute mass $m_q$,
which serves to regularize the mass (collinear) singularity.
At lowest order, both perturbative LCDAs of the fictitious $\overline{B}$ meson
are simply $\delta$ functions~\cite{Bell:2008er}:
\beq
\phi^{\rm QCD\:(0)}(x)=\delta(x-x_0),\qquad
\phi_+^{\rm HQET\:(0)}(\omega)= \delta(\omega-m_q),
\eeq
with $x_0=m_q/m_B$ and $m_B=m_b+m_q$.
From \eqref{Our:factorization:theorem}, one readily finds that
the tree-level $Z$ factor is also a simple $\delta$ function.
\beq
 Z^{(0)}(x,\omega, m_b)= \delta \left(x- {\omega\over m_b+\omega}\right).
\label{Z0:expression}
\eeq
Reassuringly, the perturbative $Z$ factor is absent of the IR scale $m_q$.
The $\delta$ function guarantees that the $\omega$ with support $(0,\infty)$
is monotonically mapped onto $x$ with support $(0,1)$.

By solving \eqref{Our:factorization:theorem} to next-to-leading order,
we then obtain the $Z$ factor of order $\alpha_s$:
\bqa
Z^{(1)}(x,\omega, m_b;\mu_Q, \mu_H) & = & \phi^{\rm QCD\:(1)}(x,\mu_Q)\Big\vert_{m_q\to \omega}-
  {m_b\over (1-x)^2} \phi_+^{\rm HQET\:(1)} \left({m_b x\over 1-x},\mu_H\right)\bigg\vert_{m_q\to \omega}
\nn\\
&-& \left(3\ln\frac{\mu_H}{m_b}+2\right)Z^{(0)}(x,\omega,m_b).
\label{Z1:expression}
\eqa
By construction, the $Z$ factor automatically obeys the ERBL equation and
LN equation.

Plugging the explicit order-$\alpha_s$ expressions for two LCDAs~\cite{Bell:2008er} into
\eqref{Z1:expression}, we end up with
\begin{align}
\label{Z1:explicit:expression}
&Z^{(1)}(x,\omega, m_b;\mu_Q,\mu_H)
\\
&= 2\left\{\left(\ln \frac{\mu_Q^{2}}{(m_b+\omega)^{2}
\left(x_{\omega}-x\right)^{2}}-1\right)\left[\left(1+\frac{1}{x_{\omega}-x}\right) \frac{x}{x_{\omega}} \theta\left(x_{\omega}-x\right)+\left(\begin{array}{c}{x \leftrightarrow \overline{x}}
\\
 {x_{\omega} \leftrightarrow \overline{x}_{\omega}}\end{array}\right)\right]\right\}_{[x]+}
\nn\\
&+4\left\{\frac{x(1-x)}{\left(x-x_{\omega}\right)^{2}}\right\}_{[x]++}+2 \delta^{\prime}\left(x-x_{\omega}\right)\left(2 x_{\omega}\left(1-x_{\omega}\right) \ln \frac{x_{\omega}}{1-x_{\omega}}+2 x_{\omega}-1\right)
\nn\\
&-\omega_x\frac{d\omega_x}{dx}\left\{2\left[\left(\ln \left[\frac{\mu_H^{2}}{(\omega_x-\omega)^{2}}\right]-1\right)\left(\frac{\theta(\omega-\omega_x)}{\omega(\omega-\omega_x)}+
\frac{\theta(\omega_x-\omega)}{\omega_x(\omega_x-\omega)}\right)\right]_{[\omega]+}+\frac{4 \theta(\omega_x-2 \omega)}{(\omega_x-\omega)^{2}}\right.
\nn\\
 &\left.+4\left[\frac{\theta(2 \omega-\omega_x)}{(\omega_x-\omega)^{2}}\right]_{[\omega]++} -\frac{\delta(\omega_x\!-\!\omega)}{\omega}\!\left(\frac{1}{2} \ln ^{2}
 \frac{\mu_H^{2}}{\omega^{2}}\!-\!\ln \frac{\mu_H^{2}}{\omega^{2}}\!+\!\frac{3 \pi^{2}}{4}\!+\!2\right)\! \right\}- \left(\!3\ln\frac{\mu_H}{m_b}\!+\!2\!\right)\!\delta(x-x_\omega),\nn
\end{align}
where for brevity we have introduced the shorthands
\beq
x_\omega\equiv\frac{\omega}{m_b+\omega},\qquad \omega_x=\frac{m_b x}{1-x}.
\eeq
The ``+'' and ``++'' functions are understood in the distributive sense,
whose exact definition can be found
in \cite{Bell:2008er}. The subscript $[x/\omega]$
enforces whether to convolute the plus function with a test function over $x$ or $\omega$.
It is reassuring that $\mu_Q$ and $\mu_H$ dependence of $Z^{(1)}$ in
\eqref{Z1:explicit:expression} are explicitly compatible  with
the evolutions equations in\eqref{ERBL:evolution:eq} and \eqref{LN:evolution:eq}.

Our refactorization program has obvious strength to optimize
the theoretical predictions for hard exclusive $B$ production processes.
Plugging \eqref{Our:factorization:theorem} into \eqref{QCD:factorization:theorem},
one obtains
\bseq
\begin{align}
& {\cal M} = \int_0^\infty\!\! d\omega\, {\mathcal T}^{\rm expd}(\omega, Q/m_b; \mu_H) \,
\Phi^{\rm HQET}_+ (\omega,\mu_H)
+{\cal O}(m_b/Q, 1/m_b),
\label{Expanded:HQET:factorization}
\\
& {\mathcal T}^{\rm expd}(\omega, Q/m_b; \mu_H)  = \int_0^1\!\! dx\, T(x,\mu_Q)\, Z(x,\omega, m_b;\mu_Q, \mu_H).
\label{Exapnded:hard:kernel:HQET:fac}
\end{align}
\label{New:HQET:factorization:formula}
\eseq

Equation \eqref{New:HQET:factorization:formula} is the desired factorization formula that
merges the virtues of both collinear and NRQCD factorization approaches,
which is assumed to yield the most optimized prediction for hard exclusive $B$ production.
As indicated in \eqref{Exapnded:hard:kernel:HQET:fac}, a more effective way of organizing calculation
is to first utilize the existing knowledge on the hard-scattering kernel $T(x)$  (typically proportional to $1/x$)
in collinear factorization. Since the quark mass has been dropped,
there is no difference for the hard-scattering kernel between $B$ and $\pi$ production.
One then employ \eqref{Exapnded:hard:kernel:HQET:fac} to obtain an effective
hard-scattering kernel ${\mathcal T}^{\rm expd}$
In accordance with the HQET factorization \eqref{Expanded:HQET:factorization},
one can convolve this effective ${\mathcal T}^{\rm expd}$ with $\Phi^{\rm HQET}_+ $ to generate
ultimate predictions.
Notice that the ${\mathcal T}^{\rm expd}$ is not identical with ${\mathcal T}$ that arises from
the literal fixed-order calculation in HQET factorization \eqref{HQET:factorization:theorem}.
Nevertheless, ${\mathcal T}^{\rm expd}$ amounts to expanding ${\mathcal T}$
to lowest order in $m_b/Q$. Consequently, ${\mathcal T}^{\rm expd}$ can only depend on
$Q/m_b$ logarithmically.
We have explicitly verified that, by invoking \eqref{New:HQET:factorization:formula},
our ${\mathcal T}^{\rm expd}$ indeed coincides with the expanded
hard kernel ${\mathcal T}$ through order $\alpha_s$ for the $W\to B\gamma$ process~\cite{Ishaq:2019zki}.

Another remarkable merit of \eqref{New:HQET:factorization:formula} is to expedite the
resummation of the collinear logarithm $\alpha_s \ln{Q\over m_b}$ to all orders.
Since the $\mu_Q$ dependence of the $Z$ function is governed by the ERBL equation,
we may follow the recipe outlined in \cite{Jia:2008ep} that also employs the ERBL equation to
resum the leading collinear logarithm
for exclusive quarkonium production, to recast \eqref{Exapnded:hard:kernel:HQET:fac} into
\begin{align}
& {\mathcal T}^{\rm expd}_{\rm LL}(\omega, Q/m_b; \mu_H)  =
\int_0^1\!\! dx\, T(x,Q)\, Z(x,\omega, m_b; Q, \mu_H)
\nn\\
&= \int_0^1\!\! dx\, T^{(0)}(x)\, \exp\left[\kappa C_F V_0*\right] Z^{(0)}(x,\omega, m_b),
\label{Leading:Log:Expanded:hard:kernel}
\end{align}
where in the second equation, we have substituted the schematic solution of the ERBL equation for
$Z(x,\omega, m_b; \mu_Q=Q)$, which is evolved from the initial IR scale at $\mu_Q\sim m_b$ to the
UV scale $\mu_Q \sim Q$. The meaning of the ``$*$''-operation will
become self-evident in below. $\kappa$ is defined by
\beq
\kappa\equiv {2\over\beta_0} \ln {\alpha_s(m^2_b)\over \alpha_s(Q^2)}\approx {\alpha_s(Q^2)\over 2\pi}\ln {Q^2\over m_b^2}
+\beta_0^2 {\alpha^2_s(Q^2)\over (4\pi)^2}\ln {Q^2\over m_b^2}+\cdots,
\label{def:kappa}
\eeq
where $\beta_0= {11\over 3}N_c-{2\over 3}n_f$ is the one-loop QCD $\beta$ function,
and $n_f=5$ is the number of active quark flavors.

Equation \eqref{Leading:Log:Expanded:hard:kernel} can be expanded iteratively~\cite{Jia:2008ep},
\begin{align}
& {\mathcal T}^{\rm expd}_{\rm LL}(\omega, Q/m_b;\mu_H)  = \int_0^1\!\! dx\,
T^{(0)}(x) Z^{(0)}(x,\omega, m_b)
\nn\\
& + \kappa C_F \int_0^1\!\! dx\,\int_0^1 \!\! dy\, T^{(0)}(x) V_0(x,y) Z^{(0)}(y,\omega,m_b)
\label{LL:truncated:Hard:Kernel:NNLO}\\
& + {\kappa^2 C^2_F\over 2} \int_0^1\!\! dx\,\int_0^1\!\! dy \,\int_0^1\!\! dz\,
T^{(0)}(x)V_0(x,y) V_0(y,z) Z^{(0)}(z,\omega, m_b)+\cdots.
\nn
\end{align}

For a leading-twist (helicity-conserving) process, one typically bears $T^{(0)}(x)\propto 1/x$.
Substituting this knowledge and \eqref{Z0:expression} into
\eqref{LL:truncated:Hard:Kernel:NNLO},
One can choose the order of multiple integration from the left to right, and leaves
the integration over $Z^{(0)}$ in the last step.
One then identifies the leading collinear logarithms at arbitrarily
prescribed perturbative order.
For example, at order $\alpha_s$, we obtain
\beq
{\mathcal T}^{\rm expd (1)}_{\rm LL} (\omega) \propto {m_b\over \omega}{\alpha_s C_F\over 4\pi}
\left(3+2\ln {\omega\over m_b}\right)\ln {Q^2\over m_b^2},
\eeq
which indeed coincides with the expanded order-$\alpha_s$ hard kernel
for the process $W\to B+\gamma$ in HQET factorization~\cite{Ishaq:2019zki}.
In a sense, this is similar to invoke the refactorization approach to recover the
NRQCD short-distance coefficients associated with exclusive $B_c$ production
through order $\alpha_s$,
when expanded to the leading order in $m_b/Q$~\cite{Jia:2010fw,Feng:2019meh}.

In summary, in this work we have unravelled a novel factorization formula \eqref{Our:factorization:theorem},
which connects two kinds of important LCDAs for heavy-flavor mesons through a perturbatively
calculable coefficient function. This perturbative function has been determined through order $\alpha_s$.
The physics underlying this factorization theorem looks quite lucid,
just because HQET shares the identical IR behavior as QCD.
It may look somewhat surprising why such a simple factorization formula has not
been discovered until now.
Based on this refactorization picture, we have devised a master formula,
\eqref{New:HQET:factorization:formula}, tailored for tackling hard exclusive $B$ production processes.
This factorization formula has inherited the virtues of both collinear and HQET factorization approaches,
which is believed to generate the most optimized theoretical predictions.
As a remarkable merit, this master formula also enables us to effectively
resum large logarithms of type $\ln Q/m_b$ and $\ln m_b/\Lambda_{\rm QCD}$
in a controlled manner.

\begin{acknowledgments}
The work of S.~I. and Y.~J. is supported in part by the National Natural Science Foundation of China
under Grants No.~11875263,
No.~11621131001 (CRC110 by DFG and NSFC). S.~I. also wishes to acknowledge the financial support
from the CAS-TWAS President's Fellowship Programme.
The work of X.-N.~X. is supported by the Deutsche Forschungsgemeinschaft (Sino-German CRC 110).
The work of D.-S.~Y. is supported in part by the National Natural Science Foundation of China under Grants No.~11275263 and 11635009.
\end{acknowledgments}


\begin{thebibliography}{99}


\bibitem{Brodsky:1989pv}
  S.~J.~Brodsky and G.~P.~Lepage,
  Adv.\ Ser.\ Direct.\ High Energy Phys.\  {\bf 5}, 93 (1989).

\bibitem{Lepage:1980fj}
  G.~P.~Lepage and S.~J.~Brodsky,
  Phys.\ Rev.\ D {\bf 22}, 2157 (1980).

\bibitem{Chernyak:1983ej}
  V.~L.~Chernyak and A.~R.~Zhitnitsky,
  Phys.\ Rept.\  {\bf 112}, 173 (1984).


\bibitem{Lepage:1979zb}
  G.~P.~Lepage and S.~J.~Brodsky,
  Phys.\ Lett.\ B {\bf 87}, 359 (1979).

\bibitem{Efremov:1979qk}
  A.~V.~Efremov and A.~V.~Radyushkin,
  Phys.\ Lett.\ B {\bf 94}, 245 (1980).


\bibitem{KPY}
G.~P.~Korchemsky, D.~Pirjol and T.~M.~Yan,
Phys.\ Rev.\ D {\bf 61} (2000) 114510.




\bibitem{DS03}
S.~Descotes-Genon and C.~T.~Sachrajda,
Nucl.\ Phys.\ B {\bf 650} (2003) 356.

\bibitem{proofsBgamma}
E.~Lunghi, D.~Pirjol and D.~Wyler,
Nucl.\ Phys.\ B {\bf 649} (2003) 349;\\
%
S.~W.~Bosch, R.~J.~Hill, B.~O.~Lange and M.~Neubert,
Phys.\ Rev.\ D {\bf 67} (2003) 094014.


\bibitem{Beneke:1999br}
  M.~Beneke, G.~Buchalla, M.~Neubert and C.~T.~Sachrajda,
  Phys.\ Rev.\ Lett.\  {\bf 83}, 1914 (1999)

\bibitem{Beneke:2000ry}
  M.~Beneke, G.~Buchalla, M.~Neubert and C.~T.~Sachrajda,
  Nucl.\ Phys.\ B {\bf 591}, 313 (2000)


\bibitem{Grozin:1996pq}
A.~G.~Grozin and M.~Neubert,
Phys.\ Rev.\ D {\bf 55} (1997) 272


\bibitem{Beneke:2000wa}
  M.~Beneke and T.~Feldmann,
  Nucl.\ Phys.\ B {\bf 592}, 3 (2001)

\bibitem{Lee:2005gza}
  S.~J.~Lee and M.~Neubert,
  Phys.\ Rev.\ D {\bf 72}, 094028 (2005)

\bibitem{Lange:2003ff}
  B.~O.~Lange and M.~Neubert,
  Phys.\ Rev.\ Lett.\  {\bf 91}, 102001 (2003)

\bibitem{Braaten:2001bf}
  E.~Braaten, Y.~Jia and T.~Mehen,
  Phys.\ Rev.\ D {\bf 66}, 034003 (2002)


\bibitem{Braaten:2001uu}
  E.~Braaten, Y.~Jia and T.~Mehen,
  Phys.\ Rev.\ D {\bf 66}, 014003 (2002)

\bibitem{Braaten:2002yt}
  E.~Braaten, Y.~Jia and T.~Mehen,
  Phys.\ Rev.\ Lett.\  {\bf 89}, 122002 (2002)

\bibitem{Ishaq:2019zki}
  S.~Ishaq, Y.~Jia, X.~Xiong and D.~S.~Yang,
  arXiv:1903.12627 [hep-ph].


\bibitem{Grossmann:2015lea}
  Y.~Grossman, M.~K\"{o}nig and M.~Neubert,
  JHEP {\bf 1504}, 101 (2015)


\bibitem{Eichten:1989zv}
  E.~Eichten and B.~R.~Hill,
  Phys.\ Lett.\ B {\bf 234} (1990) 511.

\bibitem{Neubert:1993mb}
  M.~Neubert,
  Phys.\ Rept.\  {\bf 245} (1994) 259

\bibitem{Beneke:1997zp}
  M.~Beneke and V.~A.~Smirnov,
  Nucl.\ Phys.\ B {\bf 522}, 321 (1998)

\bibitem{Ma:2006hc}
  J.~P.~Ma and Z.~G.~Si,
  Phys.\ Lett.\ B {\bf 647}, 419 (2007)


\bibitem{Xu:2016dgp}
  J.~Xu and D.~Yang,
  JHEP {\bf 1607}, 098 (2016)

\bibitem{Ji:2013dva}
  X.~Ji,
  Phys.\ Rev.\ Lett.\  {\bf 110}, 262002 (2013)

\bibitem{Xiong:2013bka}
  X.~Xiong, X.~Ji, J.~H.~Zhang and Y.~Zhao,
  Phys.\ Rev.\ D {\bf 90}, no. 1, 014051 (2014)

\bibitem{Bell:2008er}
  G.~Bell and T.~Feldmann,
  JHEP {\bf 0804} (2008) 061

\bibitem{Jia:2008ep}
  Y.~Jia and D.~Yang,
  Nucl.\ Phys.\ B {\bf 814}, 217 (2009)


\bibitem{Jia:2010fw}
  Y.~Jia, J.~X.~Wang and D.~Yang,
  JHEP {\bf 1110}, 105 (2011)

\bibitem{Feng:2019meh}
  F.~Feng, Y.~Jia and W.~L.~Sang,
  arXiv:1902.11288 [hep-ph].


\end{thebibliography}
\end{document}